\documentclass[twocolumn,showpacs,preprintnumbers]{revtex4}
\usepackage{amsmath,amssymb}
\usepackage{graphicx}
\usepackage{dcolumn}
\usepackage{bm}

\begin{document}

\title{Creation of a Photonic Time-bin Qubit via
 Parametric Interaction of Photons in a Driven Resonant Medium}
\author{N.Sisakyan}
\author{Yu.Malakyan}
\email{yumal@ipr.sci.am}

\affiliation{Institute for Physical Research, Armenian National Academy of Sciences,
Ashtarak-2, 378410, Armenia }
\date{\today }

\begin{abstract}
A novel method of preparing a single photon in
temporally-delocalized entangled modes is proposed and analyzed.
We show that two single-photon pulses propagating in a driven
nonabsorbing medium with different group velocities are temporally
split under parametric interaction into well-separated pulses. As
a consequence, the single-photon "time-bin-entangled" states are
generated with a programmable entanglement, which is easily
controlled by driving field intensity. The experimental study of
nonclassical features and nonlocality in generated states by means
of balanced homodyne tomography is discussed.
\end{abstract}

\maketitle




\section{\protect\normalsize INTRODUCTION}
Entanglement and nonlocal correlations, besides their fundamental
importance in the modern interpretation of quantum phenomena
\cite{ein, bell}, are the basic concepts for realization of
quantum information procedures \cite{ben}. The entanglement
between matter and light states is an essential element of quantum
repeaters \cite{breig}, the intermediate memory nodes in quantum
communication network aimed at preventing the photon attenuation
over long distances. The two-photon entanglement is a crucial
ingredient for quantum cryptography \cite{ekert, tittel}, quantum
teleportation \cite{bouw, furus, marc}, and entanglement swapping
\cite{zuk, pan}, which have been successfully realized during the
last decade by utilizing two approaches, one based on continuous
quadrature variables and the other using the polarization
variables of quantized electromagnetic field \cite{braun}. An
essential step has been recently made in this direction by
implementing robust sources producing the pairs of photons which
are entangled in well-separated temporal modes (time-bins)
\cite{brend}. It has been shown \cite{brend, ried} that this type
of entanglement, in contrast to other ones, can be transferred
over significantly large distances without appreciable losses,
thus being much preferable for long-distance applications. From
the fundamental viewpoint, of special interest is a single photon
delocalized into two distinct spatial \cite{knill} or temporal
modes. In the last decade, the concept of single particle
entanglement has been an object of intensive debates \cite {hardy,
peres, green, van}. Although, there was some criticism in
literature \cite {green} concerning whether a single degree of
freedom can be entangled with itself, it is now well recognized
that a single photon state delocalized spatially or temporally in
the two modes is entangled and nonlocal \cite {hardy, peres, van}.
Moreover, one-particle entangled qubit has been used to develop
and further study of quantum cryptography \cite {jlee}, quantum
computing with linear optics \cite {knill} or teleportation \cite
{villas, hlee, giac}. The robust criteria \cite {banas, hill} are
now available for verification of entanglement and nonlocality of
a correlated two-mode quantum state of light via testing the
Bell's inequality that has been recently realized experimentally
\cite{babi,zava} by performing the homodyne detection of
delocalized single-photon Fock states and reconstructing the
corresponding Wigner function from homodyne data (see also
\cite{lvov}).

Two approaches have been hitherto developed for preparation of a
single-photon in two distinct temporal modes. In first one a
time-bin qubit is created with use of linear optics by passing a
short pulse from a spontaneous parametric down conversion (SPDC)
source through Mach-Zehnder interferometer with different-length
arms \cite{brend}. The second approach is based on conditional
measurement on quantum system of entangled signal-idler pairs
generated via SPDC of two consecutive pump pulses in a nonlinear
crystal, when a detection of one idler photon tightly projects the
signal field into
a single-photon state coherently delocalized over two temporal modes \cite%
{zava}. However, the both methods are confronted with severe
challenges. The main limitation is that the light emitted via SPDC
has too broad linewidth ($\sim$ 10nm) and low spectral brightness
to be able to excite atomic species. Additionally, due to short
coherence time ($\sim$ femtosecond) the photon waveforms are not
or hardly resolvable by existing photodetectors, as well as their
coherence length is small for long distance quantum communication.

In this paper a novel method free from the above drawbacks is
discussed for dynamical preparation of photonic time-bin qubit
\begin{equation}
\mid \psi \rangle = r_{1} \mid 1\rangle_{t} \mid 0\rangle_{t+\tau}
+ r_{2} \mid 0\rangle_{t} \mid 1\rangle_{t+\tau}
\end{equation}
where $\mid 0\rangle_{t}$ and $\mid 1\rangle_{t}$ denote Fock
states with zero and one photon, respectively, at the time $t$ and
$\mid r_{1}\mid ^{2} +\mid r_{2} \mid^{2}=1$. The basic idea is to
create a parametric interaction between two single-photon pulses,
which
\begin{figure}[b]
\rotatebox{0}{\includegraphics* [scale = 0.5]{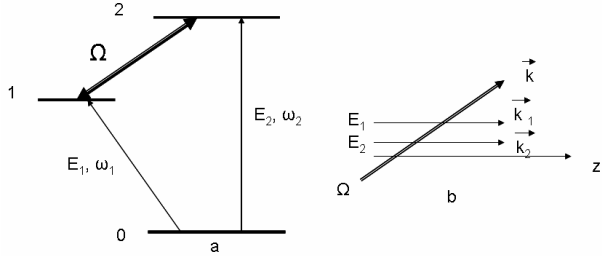}} \caption{(a)
Level scheme of atoms interacting with quantum fields $E_{1,2}$
and classical rf driving field of Rabi frequency $\Omega .$ (b)
Geometry of fields propagation.}
\end{figure}
propagate in a resonantly driven medium without absorption and at
low, but different, group velocities. Then, due to the cyclic
parametric conversion of the fields and the group delay, each
pulse experiences a temporal splitting into well-separated
subpulses. Moreover, since the process is completely coherent, at
the output of the medium the time-delocalized and entangled
single-photon state is formed. One obvious limitation of this
resonant process is that it is effective in a relatively narrow
frequency range associated with specific atoms. We note, however,
that recently a source of narrow-bandwidth, frequency tunable
single photons with properties allowing exciting the narrow atomic
resonances has been created \cite{chou, eis}. Thus, our mechanism
is a robust source for temporally entangled narrow-bandwidth
single-photons. Another important advantage is a generation in a
simple manner of any desired entanglement by controlling the
driving field intensity. In section II we describe a three-level
model parametric interaction between two quantum fields
propagating in a driven medium under the conditions of
electromagnetically induced transparency (EIT). Then, in realistic
approximations, we obtain an analytical solution for the field
operators and calculate the output intensities of the fields
showing the splitting of an initial single-photon pulse into two
well separated temporal modes. In section III we analyze the
entanglement characterizing our single-photon state by employing
the Bell's inequality proposed by Banaszek and Wodkiewicz \cite
{banas} and show an unambiguous correspondence of this inequality
violation to the degree of two-mode single-photon entanglement.
Finally, in section IV we summarize our conclusions.

\section{\protect\normalsize THREE-LEVEL MODEL OF PHOTON PARAMETRIC INTERACTION}

We consider an ensemble of cold atoms with level configuration
depicted in Fig.1. Two quantum fields
\begin{equation*}
E_{1,2}(z,t)=\sqrt{\frac{\hbar \omega _{1,2}}{2{\varepsilon }_{0}V}}%
\hat{\mathcal E}_{1,2}(z,t)\exp [i(k_{1,2}z-\omega _{1,2}t)]+h.c.
\end{equation*}
co-propagate along the $z$ axis and interact with the atoms on the
transitions $0 \rightarrow 1$ and $0 \rightarrow 2$, respectively,
while the electric-dipole forbidden transition $1 \rightarrow 2$
is driven by a classical and constant radio-frequency (rf) field
with real Rabi frequency $\Omega $ inducing a magnetic dipole or an
electric quadrupole transition between the two upper levels, and
$V$ is the quantization volume taken to be equal to interaction
volume. The electric fields are expressed in terms of the
operators $\hat{\mathcal E}_{i}(z,t)$ obeying the commutation
relations (see Appendix A)
\begin{equation}
\lbrack \hat{\mathcal E}_{i}(z,t),\hat{\mathcal
E}_{j}^{+}(z,t^{\prime})]=\frac{L}{c}\delta _{ij}\delta
(t-t^{\prime})
\end{equation}%
where $L$ is the length of the medium. We describe the latter using atomic
operators $\hat{\sigma}_{\alpha \beta }(z,t)$ $=\frac{1}{N_{z}}\underset{i=1}%
{\overset{N_{z}}{\sum }}\mid \alpha $ $\rangle _{i}\langle \beta
\mid $ averaged over the volume containing many atoms
$N_{z}=\frac{N}{L}dz\gg 1$ around position $z$, where $N$ is the
total number of atoms. In the rotating wave picture the
interaction Hamiltonian is given by
\begin{equation*}
H=-\hbar \frac{N}{L}\int\limits_{0}^{L}dz[g_{1}\hat{\mathcal E}_{1}\hat{\sigma}%
_{10}e^{ik_{1}z}+g_{2}\hat{\mathcal E}_{2}\hat{\sigma}_{20}e^{ik_{2}z}+\Omega \hat{%
\sigma}_{21}e^{ik_{\parallel }z}
\end{equation*}
\begin{equation}
+h.c.]
\end{equation}
Here $k_{\parallel }=\vec{k}_{d}\hat{e}_{z}$ \ is the projection
of the wave-vector of the driving field on the $z$ axis,
$g_{\alpha }=\mu _{\alpha 0}\sqrt{{ \omega
_{i}}/(2{\hbar\varepsilon }_{0}V})$ is the atom-field coupling
constants with $\mu _{\alpha \beta }$ being the dipole matrix
element of the atomic transition $\alpha \rightarrow \beta $. For
simplicity, we discuss the case of exactly resonant interaction
with all fields and, therefore, put in Eq.(3) the frequency
detunings equal to zero, neglecting so the Doppler broadening,
which in a cold atomic sample is smaller than all relaxation
rates. Then, using the slowly varying envelope approximation, the
propagation equations for the quantum field operators take the
form:

\begin{equation}
\left( \frac{\partial }{\partial z}+\frac{1}{c}\frac{\partial }{\partial t}%
\right) \hat{\mathcal E}_{1}(z,t)=ig_{1}\frac{N}{c}\hat{\sigma}_{01}e^{-ik_{1}z}+\hat{F%
}_{1}
\end{equation}%
\begin{equation}
\left( \frac{\partial }{\partial z}+\frac{1}{c}\frac{\partial }{\partial t}%
\right) \hat{\mathcal E}_{2}(z,t)=ig_{2}\frac{N}{c}\hat{\sigma}_{02}e^{-ik_{2}z}+\hat{F%
}_{2}
\end{equation}%
where $\hat{F}_{i}(z,t)$ are the commutator preserving Langevin
operators, whose explicit form is given below.

In the weak-field (single-photon) limit, the equations for atomic coherences $%
\hat{\rho}_{0i}=\hat{\sigma}_{0i}e^{-ik_{i}z},\ \ i=1,2~\ \ \ $and$\ \ \ \ \
\hat{\rho}_{12}=\hat{\sigma}_{12}e^{-i(k_{2}-k_{1})z}\ $ are treated
perturbatively in $\hat{\mathcal E}_{1,2}$. In the first order only $\langle \hat\sigma_{00}\rangle\simeq 1$ is different from zero and for these equations we
obtain:
\begin{equation}
\frac{\partial }{\partial t}\hat{\rho}_{01}=-\Gamma _{1}\hat{\rho}%
_{01}+ig_{1}\hat{\mathcal E}_{1}\hat{\sigma}_{00}+i\Omega \hat{\rho}%
_{02}e^{i\bigtriangleup kz}-ig_{2}\hat{\mathcal
E}_{2}\hat{\rho}_{21}
\end{equation}%
\begin{equation}
\frac{\partial }{\partial t}\hat{\rho}_{02}=-\Gamma _{2}\hat{\rho}%
_{02}+ig_{2}\hat{\mathcal E}_{2}\hat{\sigma}_{00}+i\Omega \hat{\rho}%
_{01}e^{-i\bigtriangleup kz}-ig_{1}\hat{\mathcal
E}_{1}\hat{\rho}_{12}
\end{equation}%
\begin{equation}
\frac{\partial }{\partial t}\hat{\rho}_{12}=-\Gamma _{12}\hat{%
\rho}_{12}-ig_{1}\hat{\mathcal E}_{1}^{\ast }\hat{\rho}_{02}+ig_{2}\hat{\mathcal E}_{2}\hat{\rho}%
_{10}
\end{equation}%
Here $\Delta k=k_{2}-k_{1}-k_{\parallel }$ is the wave-vector
mismatch and $\Gamma _{1,2}$ and $\Gamma _{12}$ are the transverse
relaxation rates involving, apart from natural decay rates $\gamma
_{1,2}$ of the excited states 1 and 2, the dephasing rates in
corresponding transitions. The latter are caused by atomic
collisions and escape of atoms from the laser beam. However, in
the ensemble of cold atoms the both effects are negligibly small
compared to $\gamma _{1,2}$, so that $\Gamma
_{1,2}$=$\gamma_{1,2}/2$ and $\Gamma
_{12}$=$(\gamma_{1}+\gamma_{2})/2$.

Further, we assume that the phase-matching condition $\Delta k=0$
is fulfilled in the medium. Then, the solution to Eqs.(6-8) to the
first order in $\hat{\mathcal E}_{1,2}$ is readily found to be
\begin{equation}
\hat{\rho}_{01} = i \frac{\Gamma }{D}g_{1}\hat{\mathcal
E}_{1}+i\frac{\Omega ^{2}-\Gamma
^{2}}{D^{2}}g_{1}\frac{\partial }{\partial t}\hat{\mathcal E}_{1}-\frac{\Omega }{D}%
g_{2}\hat{\mathcal E}_{2}+\frac{2\Gamma \Omega }{D^{2}}g_{2}\frac{\partial }{\partial t%
}\hat{\mathcal E}_{2},
\end{equation}%
\begin{equation}
\hat{\rho}_{02}=\hat{\rho}_{01}(1\leftrightarrow 2),\ \ \ \text{ }D=\Omega
^{2}+\Gamma ^{2}.
\end{equation}%
where, for simplicity, the optical decay rates are taken to be the same: $\Gamma _{1}=\Gamma _{2}=\Gamma$. The first terms in right hand
side (RHS) of Eqs.(9,10) are responsible for linear absorption of
quantum fields and define the field absorption coefficients
$\kappa_{i}={g_{i}^{2}\Gamma N}/{c\Omega ^{2}}$ upon substituting
these expressions into Eqs.(4,5). Here the
condition of electromagnetically induced transparency (EIT, refs. \cite%
{harris,mfleis}) $\Omega \gg \Gamma _{1,2}$ is assumed to be
satisfied for both transitions coupled to the weak-fields. Note
that the three-level configurations 0-2-1 and 0-1-2 form the
$\Lambda$- and ladder EIT-systems, respectively, with the same
decoherence time $\Gamma^{-1}$ \cite{banac}. The second terms in
RHS of Eqs.(9,10) represent the dispersion contribution to the
group velocities of the pulses, while the two rest terms describe
the parametric interaction between the fields. We require that the
photon absorption be strongly reduced
by imposing the condition $\kappa_{i}L\ll 1.$ Another limitation follows from $%
\Delta \omega _{EIT}T\geq 1$ indicating that the initial spectrum
of quantum fields is contained within the EIT window $\Delta
\omega _{EIT}={\Omega ^{2}}/({\Gamma \sqrt{\alpha }})$
\cite{fleis}, where $T$ is a duration of weak-field pulses,
$\alpha =\mathcal{N}\sigma L$ is the optical depth, $\sigma
=\frac{3}{4\pi }\lambda ^{2}$ is the resonant absorption
cross-section, and $\mathcal{N}$ is the atomic number density.
Finally, the length of the pulses has to fit the length of the medium: $Tv_{i}<L$ with $
v_{i}={c\Omega ^{2}}/{g_{i}^{2}N}$ being the group velocity of the $i$%
-th field. Taking into account that $\kappa_{i}L\sim \Gamma
^{2}\alpha /\Omega ^{2}$, this set of limitations yields
\begin{equation}
\frac{\Omega ^{2}}{\Gamma ^{2}}>>\alpha \text{ \ \ \ \ and \ }\frac{1}{\sqrt{%
\alpha }}\ll \frac{Tv_{i}}{_{L}}<1\text{\ \ \ }
\end{equation}%
It is worth noting that upon satisfying the conditions (11), the
dominant contribution to the parametric coupling between the
photons is the third term in RHS of Eq.(9,10), because in this
case the last term becomes strongly suppressed by the factor
$\Omega ^{2}T/\Gamma >>1.$

It is useful at this point to consider numerical estimations. The
sample is chosen to be $^{87}Rb$ vapor with the ground state
$5S_{1/2}(F_{g}=2)$ and exited states $5P_{3/2}(F_{e}=2)$,
$5P_{3/2}(F_{e}=3)$  being the atomic states $0$ and  $1,2$ in
Fig.1, respectively. Using the following parameters - light wavelength $%
\lambda \simeq 0.8\mu $m$,$ $\Gamma =2\pi \times 3$ MHz, atomic density $\mathcal{N}%
\sim 10^{12}$cm$^{-3}$ in a trap$\ $of length $L\sim 100$ $\mu $m,
$\Omega \sim 10\Gamma ,$ and the input pulse duration $T\simeq $
2$\div $3ns, we find $\alpha \simeq 16,$ $v_{2}\sim 10^{4}m/s$,
$v_{1}\sim 0.3v_{2},$ and $\kappa_{i}L\leqslant 0.1$. All of the
parameters we use in our calculations appear to be within
experimental reach, including the initial single-photon wave
packets with a duration of several nanoseconds satisfying the
narrow-line limitation discussed above.

The noise operators $F_{1,2}$ in Eqs.(4,5) have the properties
\cite{scully}
\begin{equation*}
\langle F_{i}(z,t)\rangle = \langle
F_{i}(z,t)F_{i}(z,t^{\prime})\rangle=\langle
F_{i}(z,t)F_{j}(z,t^{\prime})\rangle=0
\end{equation*}
\begin{equation*}
\langle
F_{i}(z,t)F_{j}^{+}(z,t^{\prime})\rangle=2\frac{\kappa_{i}}{c}\delta_{ij}\delta(t-t^{\prime})
\end{equation*}
showing that in the absence of photon losses the noise operators
$\hat{F}_{i}$ give no contribution. Then, in this limit the simple
propagation equations for the field operators are obtained:
\begin{equation}
\left( \frac{\partial }{\partial z}+\frac{1}{v_{1}}\frac{\partial
}{\partial t}\right) \hat{\mathcal E}_{1}(z,t)=-i\beta
\hat{\mathcal E}_{2}
\end{equation}%
\begin{equation}
\left( \frac{\partial }{\partial z}+\frac{1}{v_{2}}\frac{\partial
}{\partial t}\right) \hat{\mathcal E}_{2}(z,t)=-i\beta
\hat{\mathcal E}_{1}
\end{equation}%
where $\beta ={g_{1}g_{2}N/c\Omega }$ is the parametric coupling
constant. In Appendix A we show that these equations preserve the
commutation relations (2). Note that for the parameters above, the
parametric interaction between the photons is sufficiently strong:
$\beta L\sim 3$.

The formal solution of Eqs.(12,13) for the field operators in the region $%
0\leqslant z\leqslant L$ is written as
\begin{equation*}
\hat{\mathcal E}_{i}(z,t)=\hat{\mathcal E}_{i}(0,\ t-z/v_{i})+\int\limits_{0}^{z}dx\{\hat{\mathcal E}%
_{i}(0,\ t-z/v_{j}-\frac{\Delta v_{ji}}{v_{i}v_{j}}x)
\end{equation*}%
\begin{equation}
\times \frac{\partial J_{0}(\psi )}{\partial z}-i\beta \
\hat{\mathcal E}_{j}(0,\ t-z/v_{i}-\frac{\Delta
v_{ij}}{v_{i}v_{j}}x)\ J_{0}(\psi )\},\
\end{equation}%
where $i,j=1,2$ \ and $j\neq i.$ The Bessel function $\ J_{0}(\psi )$
depends on $z$ via $\psi =2\beta \sqrt{x(z-x)}$ , $\Delta v_{ij}=v_{i}-v_{j}$
is the difference of group velocities.

We are interested in the evolution of the input state $\mid \psi
_{in}\rangle = \mid 1_{1}\rangle \otimes \mid 0_{2}\rangle $
consisting of a single-photon wave packet at $\omega _{1}$
frequency, while  $\omega _{2}$ field is in the vacuum state. The
similar results are clearly obtained in the case of
one input photon at $\omega _{2}$ frequency. We assume that initially the $%
\omega _{1}$ pulse is localized around $z=0$ with a given temporal profile $%
f_{1}(t):$%
\begin{equation}
\langle 0\mid \hat{\mathcal E}_{1}(0,t)\mid \psi _{in}\rangle =\langle 0\mid \hat{\mathcal E}%
_{1}(0,t)\mid 1_{1}\rangle =f_{1}(t)
\end{equation}%
where $f_{1}(t)$ is normalized as $\frac{c}{L}\int\mid
f_{1}(t)\mid ^{2}dt=1$. In free space, $\hat{\mathcal
E}_{1}(z,t)=\hat{\mathcal E}_{1}(0,t-z/c)$ and we have
\begin{equation}
\langle 0\mid \hat{\mathcal E}_{1}(0,t-z/c)\mid 1_{1}\rangle
=f_{1}(t-z/c).
\end{equation}%
The intensities of the fields at any distance in the region $%
0\leqslant z\leqslant L$ are given by%
\begin{equation}
\langle I_{i}(z,t)\rangle =\mid \langle 0\mid \hat{\mathcal
E}_{i}(z,t)\mid \psi _{in}\rangle \mid ^{2}
\end{equation}%
Using Eqs.(14-16) and recalling that $\langle 0\mid \hat{\mathcal
E}_{2}(0,t)\mid \psi _{in}\rangle =0$, we calculate $\langle
I_{i}\rangle $ numerically and show in Fig.2 the output pulses at
$z=L$ for the three values of $\Omega $ and for Gaussian input (at
$z=0$) pulse $f_{1}(t)=C\exp [-2t^{2}/T^{2}]$, where $C$ is some
normalization constant. For one-photon initial state, as is the
case here, one can clearly see that the second\ field is not
practically generated, thus demonstrating that our scheme enables
to prepare a single-photon in a pure temporally-delocalized state
with an efficiency $\sim 100\%.$ Moreover, depending on the
driving field intensity, a different degree of initial pulse
splitting is attainable. It is easy to check that the total number
of photons which is determined by the areas of the corresponding
peaks is conserved upon propagation through the medium. To show
this we introduce the dimensionless operators for numbers of
photons that pass each point on z axis over the whole of time
\begin{equation}
\hat n_{i}(z)=\frac{c}{L}\int dt\hat{\mathcal
E}_{i}^{+}(z,t)\hat{\mathcal E}_{i}(z,t)
\end{equation}%
With taking into account that $\langle 0\mid \hat{\mathcal
E}_{i}(z,t\rightarrow\pm\infty)\mid \psi _{in}\rangle=0$, the
conservation low for mean photon numbers results from Eqs.(12) and
(13)
\begin{equation}
\frac{\partial }{\partial z}(n_{1}(z)+n_{2}(z)) =0
\end{equation}%
\begin{figure}[b]
\rotatebox{0}{\includegraphics* [scale = 1]{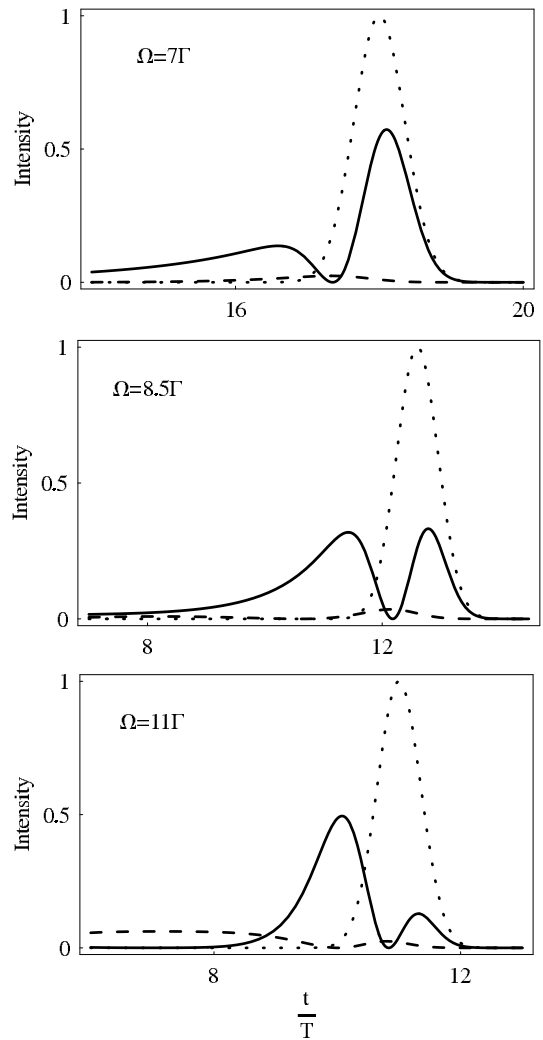}}
\caption{Temporal shapes of quantum fields ($t$ is given in units
of input pulse duration $T$) at the output of the medium $z=L$ for
three values of $\Omega $. In these figures, solid curves
represent
the $\protect\omega _{1}$ pulses, dashed curves show the $\protect%
\omega _{2}$ field generated in the medium, and dotted lines
display the initial Gaussian pulse at $\protect\omega _{1}$
frequency with $T=2$ns
propagating in the medium in the absence of parametric interaction $\protect%
\beta =0.$ For the rest of parameters see the text.}
\end{figure}
where $n_{i}(z)=\langle \psi_{in}\mid \hat n_{i}(z)\mid \psi
_{in}\rangle$. Since in our case $n_{2}(0)=0$ and $n_{2}(L)$ is
negligibly small, we have $n_{1}(L)\simeq n_{1}(0)=1$. Thus, in
the considered scheme we are able to preserve the output  $\omega
_{1}$-field in a single-photon state while modulating its
amplitude to get a desirable spatio-temporal distribution.
Further, only two well-separated output temporal modes at $\omega
_{1}$ frequency are produced. To understand the physics of this
splitting, let us discuss the structure of solution (14) for
$\hat{\mathcal E}_{i}(z,t)$ in detail. The first term in this
equation represents the $\omega_{1}$-pulse in the absence of
$\omega_{2}$ generation. In such a case, the group velocity of the
pulse is slowed down to $v_{1}<<c$ under the conditions of EIT
realized via ladder system 0-1-2. However, the input $\omega_{1}$
photon can also be converted to one photon of the $\omega_{2}$
field which is emitted on the dipole-allowed transition
$2\rightarrow0$ and propagates in the medium at a group velocity
$v_{2}<<c$ established under EIT in the $\Lambda$- scheme 0-2-1.
In its turn, the $\omega_{2}$ photon is transformed back into
$\omega_{1}$ photon, which precedes the signal $\omega_{1}$ photon
owing to $v_{2}> v_{1}$. This process is described by the second
term in Eq.(14). The last term in this equation corresponds to
generation of an $\omega_{1}$ photon by an input $\omega_{2}$
photon and gives no contribution in our case. An important point
here is that an atom, being excited to the upper state 2 upon
absorbing the initial $\omega_{1}$ photon and one photon of the
drive field, can return back to the ground state in two ways: 1)
by emitting one photon of drive field and an $\omega_{1}$ photon
and 2) by emitting an $\omega_{2}$ photon on the transition
$2\rightarrow 0$. The competition between the two processes
evidently leads to a destructive interference between the modes
propagating in the $\omega_{1}$ channel with different group
velocities and ultimately gives rise to the pulse temporal
splitting. In Appendix B we show that, indeed, the contributions
of the two processes into $\omega_{1}$ photon wavefunction are of
opposite signs. The separation of $\omega_{1}$ temporal modes
depends clearly on the relative velocity of quantum fields, the
larger the ratio $v_{2}/v_{1}$, the larger the group delay and the
larger the separation of the two output $\omega _{1}$ pulses. On
the contrary, in the limit of equal group velocities
$v_{2}=v_{1}=v$ the propagating pulses experience no splitting, as
it follows from Eqs.(14), which in this case are reduced to
\begin{equation}
\hat{\mathcal E}_{i}(z,t)=\hat{\mathcal E}_{i}(0,\tau )cos(\beta
z)-i\hat{\mathcal E}_{j}(0,\tau )sin(\beta z)
\end{equation}%
where $\tau =t-z/v$, $j\neq i.$

We finish this consideration with a short remark about the
dynamics of frequency conversion from $\omega _{1}$ to  $\omega
_{2}$  and back to  $\omega _{1}$  in dependence on the travelled
distance. For equal group velocities $v_{1}=v_{2}$, a simplest
result follows from Eq.(20) with $\langle 0\mid \hat{\mathcal
E}_{2}(0,t)\mid \psi _{in}\rangle =0$. It is seen that the
complete conversion of the input $\omega _{1}$ photon to one
photon of the $\omega _{2}$ field occurs at the distance
$z_{max}(v_{1}=v_{2})= \pi/(2\beta )$, which is proportional to
$\Omega$ and thus exhibits the square-root dependence on the drive
field intensity. An essentially different picture is observed for
$v_{1}\neq v_{2}$. In this case the efficient frequency conversion
from $\omega _{1}$ to $\omega _{2}$ takes place up to a distance,
where the time delay between the signal $\omega _{1}$ and
parametrically generated $\omega _{2}$ pulses becomes comparable
to the pulse duration $T$. The numerical calculations show that,
for the parameters above, this happens roughly at
$z_{max}(v_{1}\neq v_{2})\simeq( v_{2}-v_{1})T\sim 20\mu$m. At
this point, the $\omega _{2}$ field reaches its maximal value with
conversion efficiency $\sim$ 0.13. For $z> z_{max}(v_{1}\neq
v_{2})$ the two pulses are no longer overlapped in time, and the
$\omega _{2}$ pulse is transformed to the fast $\omega _{1}$ pulse
almost completely. Only a tiny part on the leading edge of the
$\omega _{2}$ pulse leaves the medium with the group velocity
$v_{2}$. With further propagation in the medium, the energy is
merely pumped from the slow $\omega _{1}$ pulse to the fast one,
while the $\omega _{2}$ field remains negligibly small. This
occurs as long as the fast $\omega _{1}$ pulse becomes
sufficiently strong in order to generate a new $\omega _{2}$
pulse. According to our analysis, the corresponding distance is
approximately 130$\mu$m showing that in a sufficiently large
interval of propagation lengths only the $\omega _{1}$ field is
present in the medium in the form of two well separated temporal
modes. This result provides a wide choice of the length of atomic
sample that is important for further applications of the proposed
mechanism.

It must be noted that the considered system is capable of fully
entangling two single-photon pulses at different frequencies
$\omega _{1}$ and $\omega_{2}$ in the case of input state $\mid
\psi _{in}$ $\rangle =$ $\mid 1_{1}$ $\rangle \otimes \mid 1_{2}$
$\rangle $. The study of this problem is, however, beyond the
scope of the present paper and its results will be published
elsewhere. Here we note only that in this case two time-bin qubits
at $\omega _{1}$ and $\omega _{2}$ are generated, being at the
same time strongly correlated with each other. This correlation is
clearly seen from the particular result of Eq.(20).

\section{\protect\normalsize NONLOCALITY IN GENERATED STATE: ANALYSIS OF BELL'S INEQUALITY}

Now we discuss whether nonlocal correlations arise between the two
generated temporal modes of a single $\omega _{1}$-photon and how
they can be verified experimentally. We first note that the
entanglement is not apparent in the single-photon wavefunction
$\Phi_{1}(z,t)$, which is represented as a sum (see Eq.(B3)),
rather than mixture product of two orthogonal mode functions
$\Phi_{1}^{F} (z,t)$ and $\Phi_{1}^{S} (z,t)$. This does not mean,
however, that the entanglement is absent in the photon state. It
can appear in second quantization formalism, where the solution
Eqs.(14) for the field operators has been found. The point is what
representation of Fock space has to be chosen to make this effect
visible, since, as it has been shown in \cite {pawl}, the
entanglement with vacuum and nonlocality in a single-photon state
is not a property of the Fock space in general, but appears if a
specific irreducible representation is chosen, although the same
physics follows from all Fock representations in the sense that
the experimental test of entanglement and nonlocality is performed
by mapping the quantum field state into the state of the matter
(single trapped atoms or ions, atomic ensembles, quantum dots,
etc.) and the results of such measurements are the same
independent of the representation. Below we describe the output
$\omega _{1}$-photon state in terms of quantized temporal modes
and choose the relevant representation for the field quantization.

Using the definitions of photon number operators Eq.(18) and
wavefunction Eq.(B1), we obtain the $\omega_{1}$ single-photon
output state as
\begin{equation}
\mid 1_{1}\rangle_{z=L}=\frac{c}{L}\int dt
\Phi_{1}(L,t)\hat{\mathcal E}_{1}^{+}(L,t)\mid 0\rangle
\end{equation}
Normalization requires that
\begin{equation*}
\frac{c}{L}\int dt \mid \Phi_{1}(L,t)\mid^2=1
\end{equation*}
This condition, the left hand side of which coincides evidently
with the mean photon number $n_{1}(L)$, manifests along with
$n_{1}(0)=\frac{c}{L}\int dt \mid f_{1}(t)\mid^2=1$ the photon
number conservation law. Remind that for intermediate values of
$z$ ($0<z<L$) $n_{2}(z)$ is no longer zero and hence $n_{1}(z)<1$.
At these distances the incoming $\omega _{1}$ field is converted,
completely or partially, into $\omega _{2}$ optical mode that
enables frequency conversion and redistribution of quantum
information between different quantum fields. This mechanism will
be discussed elsewhere.

Let us rewrite the Eq.(21) in the form
\begin{equation}
\mid 1_{1}\rangle_{z=L}=\frac{c}{L}\int dt
(\Phi_{1}^{F}(L,t)+\Phi_{1}^{S}(L,t))\hat{\mathcal
E}_{1}^{+}(L,t)\mid 0\rangle
\end{equation}
and introduce the operators of creation of single-photon wave
packets associated with orthogonal set of mode functions
$\Phi_{1}^{i}(L,t)$ \cite {blow}, where $i$ labels the members of
the denumerably infinite set. For $i=F,S$ these operators are
given by
\begin{equation}
\hat c_{F,S}^{+}=N_{F,S}^{1/2}\int dt
\Phi_{1}^{F,S}(L,t)\hat{\mathcal E}_{1}^{+}(L,t)
\end{equation}
with the normalization constants
\begin{equation}
N_{i}=\frac {c}{L}(\int dt \mid\Phi_{1}^{i}(L,t)\mid^2)^{-1}
\end{equation}
These operators create the single-photon states in the usual way
by operation on the vacuum state $\mid 0\rangle$
\begin{equation}
\hat c_{i}^{+}\mid 0\rangle=\mid 1_{1}\rangle_{i}
\end{equation}
and have the standard boson commutation relations
\begin{equation}
[\hat c_{i},\hat c_{j}^{+}]=\delta_{ij}
\end{equation}
Note that this definition of quantum temporal modes is only
useful, if one can perform the local measurements on these modes
such that they are spacelike separated. That is why the
requirement for the modes be well separated is important. Now, for
the algebra (26) we choose the representation of infinite product
of all vacua
\begin{equation}
\mid 0\rangle=\prod_{i}\mid 0\rangle_{i}=\mid 0\rangle_{F}\mid
0\rangle_{S}\prod_{i\neq F,S} {\mid 0\rangle_{i}}.
\end{equation}
However, since in our problem we deal with two modes, while the
other modes are not occupied by the photons and, hence, are not
taken into account during the measurements, the vacuum may be
reduced to $\mid 0\rangle=\mid 0\rangle_{F}\mid 0\rangle_{S}$.
Then the single-photon state (22) can be written as
\begin{equation*}
\mid 1_{1}\rangle_{z=L}=r_{1}\hat c_{F}^{+}\mid 0\rangle_{F}\mid
0\rangle_{S}+ r_{2}\mid 0\rangle_{F}\hat c_{S}^{+}\mid
0\rangle_{S}
\end{equation*}
\begin{equation}
=r_{1}\mid 1\rangle_{F}\mid 0\rangle_{S}+r_{2}\mid
0\rangle_{F}\mid 1\rangle_{S}
\end{equation}
which is just the state (1) with $r_{1,2}=\sqrt{\frac {c}{L}\int
dt \mid\Phi_{1}^{F,S}(L,t)\mid^2}$. The remarkable property of
chosen representation (27) is that in this case the entanglement
in the single-photon state is entirely converted into nonlocal
entanglement between the atoms \cite {pawl} and, hence, the
entanglement in the field state reproduces adequately the expected
results of any measurement one may perform on atomic systems. The
amount of entanglement in the state (28) is simply
\begin{equation}
E=-[r_{1}^2\log_{2}r_{1}^2+(1-r_{1}^2)\log_{2}(1-r_{1}^2)]
\end{equation}
calculated as $E=-Tr_{S}[\rho\log_{2}\rho]$ with $\rho=Tr_{F}\mid
1_{1}\rangle\langle 1_{1}\mid$. It is easy to check that $E$ is
maximal $E_{max}=1$ for $r_{1}=\frac{1}{\sqrt 2}$.

Now we pass to discussion of nonlocal correlations in the state
(28). Note that the single-photon states are completely described
by their Wigner function, whose remarkable property is that it
takes negative values around the origin of phase space for the
complex field amplitude. The negativity of the Wigner function is
the ultimate signature of non-classical nature of these states.
Besides, the nonlocality of quantum correlations in single-photon
qubit is directly evident from the violation of Bell's inequality
formulated for two-mode Wigner function. Specifically, we will
employ the criterion proposed by Banaszek and Wodkiewicz \cite
{banas} allowing a Bell test with high levels of violation. The
ability of this approach has recently been demonstrated in the
case of SPDC temporally entangled single photon \cite{zava}. As
has been shown in Ref.\cite {banas}, the local theories impose the
bound
\begin{equation}
-2\leqslant {\mathcal B}\leqslant 2
\end{equation}
where the combination ${\mathcal B}$ has the form:
\begin{equation*}
{\mathcal B}=\frac{\pi ^{2}}{4}[W(0,0)+W(\alpha _{1},0)+W(0,\alpha
_{2})-W(\alpha _{1},\alpha _{2})]
\end{equation*}
\begin{figure}[b]
\rotatebox{0}{\includegraphics* [scale=0.8]{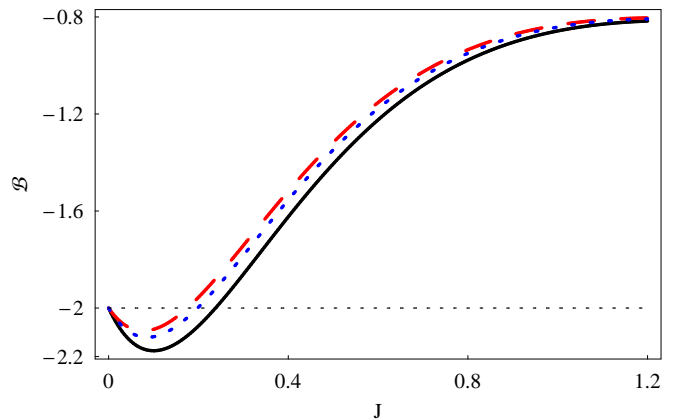}} \caption{(color
online) Plot of the combination ${\mathcal B(J)}$ (Eq.(32)) for
three values of $r_{1}$: 0.4 (dashed); $1/\sqrt{2}$ (solid) and
0.9 (dotted) corresponding to three values of $\Omega $ in Fig.2.
(from top to bottom).}
\end{figure}
Here $W(\alpha _{1},\alpha _{2})$ is the Wigner function of two
temporal modes calculated for complex amplitudes $\alpha _{i}=x_{i}+iy_{i}$ with $x_{i}$ and $y_{i}, i=1,2,%
$ being the quadratures of the $i$-th mode. The Wigner function
for the state (28) is obtained to be:
\begin{equation}
W(\alpha _{1},\alpha _{2})=\frac{4}{\pi ^{2}}[2\mid r_{1}\alpha
_{1}+r_{2} \alpha _{2}\mid ^{2}-1] e^{-2\mid \alpha _{1}\mid
^{2}-2\mid \alpha _{2}\mid ^{2}}
\end{equation}%
which is always negative $W(0,0)<0$ independent of $r_{1}$ and
$r_{2}$. The amplitudes $r_{1}$ and $r_{2}$ depend on the driving
field intensity and are calculated numerically.

Then the strongest violation of inequality (30) is achieved when
$\alpha=\alpha_{1}=\alpha_{2}$, for which case the combination
${\mathcal B}$ takes the form:
\begin{equation}
{\mathcal
B}=-1+(4J-2)e^{-2J}-[4J(r_{1}+\sqrt{1-r_{1}^2})^2-1]e^{-4J}
\end{equation}
where $J=\alpha^2$. Its behavior is plotted in Fig.3 as a function
of $J$ for the values of $r_{1}$ corresponding to three output
states of $\omega_{1}$ photon depicted in Fig.2. It is evident
that the maximal violation, which is about ten percent (2.2
compared to a classical maximum of 2), is obtained at $\alpha=0.3$
for $r_{1}=1/\sqrt{2}$, i.e. when the output temporal modes are
produced with equal probabilities. Similar to the previous works
\cite{zava,lvov}, the Wigner function (22) can be experimentally
reconstructed from the data of balanced homodyne detection, when
the $\omega_{1}$ signals at the detectors are measured at two
different times matched to the time separation between two output
pulses obtained in Fig.2. However, a rigorous experimental
demonstration of maximal violation is attainable in a
loophole-free Bell test, when both locality and detection
loopholes are closed in a single experiment \cite{loop}. The
locality-loophole can be easily avoided, if the homodyne detection
of two co-propagating temporal modes is performed by one-photon
detectors as fast as the two simultaneous measurements on each of
the modes are separated by a spacelike interval. This may easily
be achieved within our model, because the time separation between
the two temporal modes amounts to several nanoseconds. To
eliminate second, detection-efficiency loophole, we follow the
works \cite{zava}, although in these papers SPDC in nonlinear
crystal has been used for generation of time-bin qubit. To satisfy
the Banaszek-Wodkiewicz criterion one has to prepare pure
single-photon state or the "vacuum-cleaned" Wigner function. To
this end the different methods have been applied in \cite{zava}
leading to the same result and showing a strict violation of
Bell's inequality in accordance with theoretical expectations.

\section{\protect\normalsize CONCLUSIONS}

In conclusion, we have demonstrated the possibility for dynamic
preparation of a single photon in distinct temporal modes,
employing strong parametric interaction between two slow
single-photon pulses and their group delay. Disregarding the
photon losses, we have found the solution of propagation equations
for the quantum field operators depending on the propagation
distance in terms of the Bessel function. Although the losses
result in a decay of the field amplitudes, they do not prevent the
temporal splitting of quantum pulses. Moreover, since the two well
separated $\omega_{1}$ pulses undergo the same losses, the
entanglement between them is almost insensitive to losses and
easier to purify, so that the proposed scheme can be regarded as a
robust source of narrow-bandwidth single-photon qubits. We have
shown the ability of our scheme to achieve an arbitrary
entanglement between the $\omega_{1}$ temporal modes by adjusting
the driving field intensity, while the separation between the time
bins can be controlled by using the different atomic-level
configurations to obtain the different group velocities of quantum
fields. Subsequent papers will discuss a possibility for
transferring and distributing quantum information between optical
modes of different frequencies $\omega_{1}$ and $\omega_{2}$ in a
loss and decoherence-free fashion, as well as the more complicated
case of two input single-photon pulses will be analyzed and  the
results of detail numerical simulations will be presented.

\bigskip

This work was supported by the ISTC Grant No.A-1095 and INTAS
Project Ref.Nr 06-1000017-9234.

\bigskip

{\protect\normalsize \textbf{APPENDIX A: COMMUTATION RELATIONS FOR
FIELD OPERATORS}}

In this Appendix we derive the commutation relations Eq.(1) for
traveling-wave electric field operators. In free space $z\leq0$,
the field operators are given by $\hat{\mathcal
E}_{i}(z,t)=\underset{n=0} {\overset{\infty}{\sum
}}a_{i,n}e^{-i\omega_{n}(t-z/c)}$, where $a_{i,n}$ is the
annihilation operator for the $i$-th field discrete mode with a frequency $ \omega_i + \omega_n$. These
operators satisfy the standard boson commutation relations
\begin{equation}
[a_{i,n},a_{j,n^{\prime }}^{+}]=\delta_{ij}\delta_{nn^{\prime }}
\end{equation}
In the continuum limit $a_{i,n}\rightarrow (\Delta\omega)^{1/2}
a_{i}(\omega)$ and $\sum\rightarrow (1/\Delta\omega)\int d\omega$
with $\Delta\omega=2\pi c/L$, the commutation relations at z=0 are
found to be
\begin{equation}
\lbrack \hat{\mathcal E}_{i}(0,t),\hat{\mathcal
E}_{j}^{+}(0,t^{\prime })]=\frac{L}{c}\delta _{ij}\delta
(t-t^{\prime })\tag{A1}
\end{equation}%
Inside the medium, the commutation relations satisfy the equation
\begin{equation*}
\frac{\partial}{\partial z}[\hat{\mathcal
E}_{1}(z,t),\hat{\mathcal E}_{1}^{+}(z,t^{\prime
})]=-\frac{1}{v_{1}}(\frac{\partial}{\partial
t}+\frac{\partial}{\partial t^{\prime }})[\hat{\mathcal
E}_{1}(z,t),\hat{\mathcal E}_{1}^{+}(z,t^{\prime })]
\end{equation*}
\begin{equation}
-i\beta([\hat{\mathcal E}_{2}(z,t),\hat{\mathcal
E}_{1}^{+}(z,t^{\prime })]-[\hat{\mathcal
E}_{1}(z,t),\hat{\mathcal E}_{2}^{+}(z,t^{\prime })])\tag{A2}
\end{equation}%
With the use of Eqs.(14) we have
\begin{equation*}
[\hat{\mathcal E}_{2}(z,t),\hat{\mathcal E}_{1}^{+}(z,t^{\prime
})]=[\hat{\mathcal E}_{1}(z,t),\hat{\mathcal
E}_{2}^{+}(z,t^{\prime })]=0
\end{equation*}%
Recalling also that $[\hat{\mathcal E}_{1}(z,t),\hat{\mathcal
E}_{1}^{+}(z,t^{\prime })]$ is a function of time difference
$t-t^{\prime }$ and, hence,
\begin{equation*}
(\frac{\partial}{\partial t}+\frac{\partial}{\partial t^{\prime
}})[\hat{\mathcal E}_{1}(z,t),\hat{\mathcal E}_{1}^{+}(z,t^{\prime
})]=0
\end{equation*}
we obtain that the commutation relations are spatially invariant
and have the form of Eq.(2).

\bigskip

{\protect\normalsize \textbf{APPENDIX B: DESTRUCTIVE INTERFERENCE
IN PULSE TEMPORAL SPLITTING}}

The goal of this appendix is to show that the two processes
responsible for $\omega_{1}$ pulse splitting interfere
destructively. For that we calculate the $\omega_{1}$ single
photon wavefunction
\begin{equation*}
\Phi_{1}(z,t)= \langle 0\mid \hat{\mathcal E}_{1}(z,t)\mid
1_{1}\rangle = f_{1}(t-z/v_{1})+
\end{equation*}
\begin{equation}
+\int\limits_{0}^{z}dxf_{1}(t-z/v_{2}-\frac{
v_{2}-v_{1}}{v_{1}v_{2}}x)\frac{\partial J_{0}(\psi )}{\partial z}
\tag{B1}
\end{equation}%
where Eq.(16) has been used. Using instead of $x$ a new variable
$y=x/z$ and taking into account
\begin{equation*}
\frac{\partial J_{0}(\psi )}{\partial z}=-J_{1}(\psi
)\frac{\partial \psi}{\partial z}= -J_{1}(\psi )\beta\sqrt{
\frac{y}{1-y}}
\end{equation*}
where $\psi$ is now written as $\psi=2\beta z\sqrt{y(1-y})$, we
finally get
\begin{equation*}
\Phi_{1}(z,t)= f_{1}(t-z/v_{1})-
\end{equation*}
\begin{equation}
-z\beta\int\limits_{0}^{1}dy f_{1} (t-z/v_{2}-\frac{
v_{2}-v_{1}}{v_{1}v_{2}}zy)J_{1}(\psi )\sqrt{ \frac{y}{1-y}}
\tag{B2}
\end{equation}
It is easy to see that $\psi$ varies within the limits
$0<\psi<\beta z$ where, for the chosen parameters, $\beta
z\leqslant\beta L\sim 3$. Since in this interval $J_{1}(\psi )>0$,
the integrand in the second term in Eq.(A2) is positive and, thus,
the parametric regeneration of $\omega_{1}$ photon interferes
destructively with the first term $f_{1}(t-z/v_{1})$, which
describes the $\omega_{1}$ pulse slowing under the EIT in the
ladder system.

It is convenient to express $\Phi_{1}(z,t)$ as a sum of two
wavefunctions corresponding to the fast and slow
$\omega_{1}$-pulses
\begin{equation}
\Phi_{1}(z,t)= \Phi_{1}^{F}(z,t)+\Phi_{1}^{S}(z,t) \tag{B3}
\end{equation}
where $\Phi_{1}^{F,S}(z,t)$ are real functions and
$\Phi_{1}^{F}(z,t)$ is always negative. By numerical calculations
one can show that, for a given time,
\begin{equation}
\int\Phi_{1}^{F}(z,t)\Phi_{1}^{S}(z,t)dz=0 \tag{B4}
\end{equation}
and at a fixed $z$
\begin{equation}
\int\Phi_{1}^{F}(z,t)\Phi_{1}^{S}(z,t)dt=0 \tag{B5}
\end{equation}
indicating that the two temporal modes are spatially and
temporally well separated systems sharing one photon.


\bigskip

\begin{references}

\bibitem{ein} A. Einstein, B.Podolsky, and N.Rosen, Phys.Rev. {\bf 47}, 777 (1935).

\bibitem{bell} J.S.Bell, Physics {\bf 1}, 195 (1964).

\bibitem{ben} C.H.Bennett and B.D.Di Vincenzo, Nature (London) {\bf 404}, 247 (2000).

\bibitem{breig} H.-J.Briegel, W.Dur, J.I.Cirac, and P.Zoller, Phys.Rev.Lett. {\bf 81}, 5932 (1998).

\bibitem{ekert} A.Ekert, Phys.Rev.Lett. {\bf 67}, 661 (1991).

\bibitem{tittel}  W. Tittel, J. Brendel, H. Zbinden, and N. Gisin, Phys.Rev.Lett. {\bf 84}, 4737 (2000).

\bibitem{bouw} D.Bouwmeester, et al., Nature (London) {\bf 390}, 575 (1977).

\bibitem{furus} A.Furusawa, et al., Science {\bf 282}, 706 (1998).

\bibitem{marc}  I. Marcikic†, H. de Riedmatten†, W. Tittel†,‡, H. Zbinden† and N. Gisin†, Nature (London) {\bf 421}, 509 (2003).

\bibitem{zuk}  M.Zukowski, A.Zeilinger, M.A.Horne, and A.K.Ekert, Phys.Rev.Lett. {\bf 71}, 4287 (1993).

\bibitem{pan} J.-W.Pan, D.Bouwmeester, H.Weinfurter, and A.Zeilinger, Phys.Rev.Lett. {\bf 80}, 3891 (1998).

\bibitem{braun} S.L.Braunstein and P.van Loock, Rev.Mod.Phys. {\bf 77}, 513 (2005).

\bibitem{brend} J. Brendel, N. Gisin, W. Tittel, and H. Zbinden, Phys.Rev.Lett.{\bf 82}, 2594 (1999).

\bibitem{ried} H.de Riedmatten, et al., Phys.Rev.Lett. {\bf 92}, 047904 (2004).

\bibitem{knill} E. Knill, R. Laflamme, and G. J. Milburn, Nature (London) {\bf 409}, 46 (2001).

\bibitem{hardy} L.Hardy, Phys.Rev.Lett. {\bf 73}, 2279 (1994); ibid.{\bf 75}, 2065 (1995).

\bibitem{peres} A.Peres, Phys.Rev.Lett. {\bf 74}, 4571 (1995).

\bibitem{green} D.M.Greenberger, M.A.Horne, A.Zeilinger in \emph{Quantum
Interferometry,} edited by F.De Martini et al (VCH Publishers,
Weinheim, 1996), p.119.

\bibitem{van} S.J.van Enk, Phys.Rev. {\bf A72}, 064306 (2005).

\bibitem{jlee} J.W.Lee et al., Phys.Rev. {\bf A68}, 012324 (2003).

\bibitem{villas} C.J.Villas-Boas, N.G. de Almeida, and M.H.Moussa, Phys.Rev. {\bf A60}, 2759 (1999).

\bibitem{hlee} H.-W.Lee and J.Kim, Phys.Rev. {\bf A63}, 012305 (2000).

\bibitem{giac} S.Giacomini, F.Sciarrino, E.Lombardi, and F. De Martini, Phys.Rev. {\bf A66}, 030302 (2002).

\bibitem{banas} K.Banaszek and K.Wodkiewicz, Phys.Rev.Lett.{\bf 82}, 2009 (1999).

\bibitem{hill} M.Hillary and M.S.Zubairy, Phys.Rev.Lett. {\bf 96}, 050503 (2006); quant-ph/0606154.

\bibitem{babi}  S. A. Babichev, J. Appel, and A. I. Lvovsky, Phys.Rev.Lett. {\bf 92}, 193601 (2004).

\bibitem{zava}  A. Zavatta, M.D'Angelo, V.Parigi, and M.Belini, Phys.Rev.Lett. {\bf 96}, 020502 (2005); M. D`Angelo, A. Zavatta, V. Parigi, and M. Bellini, Phys.Rev. {\bf A74}, 052114 (2006).

\bibitem{lvov}  A. I. Lvovsky, et al., Phys.Rev.Lett. {\bf 87}, 050402 (2001).

\bibitem{chou} C.W. Chou, S.V. Polyakov, A. Kuzmich, and H. J. Kimble, Phys. Rev. Lett. {\bf 92}, 213601 (2004).

\bibitem{eis} M.D.Eisaman et al., Nature (London) {\bf 438}, 837 (2005).

\bibitem{harris}  S.E.Harris, Phys.Today {\bf 50}, 36 (1997).

\bibitem{mfleis}  M.Fleischhauer, A.Imamoglu, and  J.Marangos, Rev.Mod.Phys. {\bf 77}, 633 (2006).

\bibitem{banac}  Theoretical analysis of EIT in a ladder system and
comparison with experiment can be found in J.Gea-Banacloche, Y.-q.
Li, S-z. Jin, and M.Xiao, Phys.Rev. {\bf A51}, 576 (1995).

\bibitem{scully}  M.O.Scully and M.S.Zubairy, \emph{Quantum Optics}(Cambridge University Press, Cambridge, UK,1997).

\bibitem{fleis} M. Fleischhauer and M. D. Lukin, Phys. Rev. Lett. {\bf 84}, 5094 (2000); Phys. Rev. {\bf A65}, 022314 (2002).

\bibitem{pawl} M.Pawlowski and M.Czachor, Phys.Rev. {\bf A73}, 042111 (2006).

\bibitem{blow} K.Blow, R.Loudon,and S.Phoenix, and T.Shepherd, Phys.Rev. {\bf A42}, 4102 (1990).

\bibitem{loop}  For discussion of loophole-free test of Bell's inequalities see
C.Simon and W.T.M.Irvine, Phys.Rev.Lett. {\bf 91}, 110405 (2003);
R.Garcia-Patron et al., ibid. {\bf 93}, 130409 (2004) and
references therein.

\end{references}
\end{document}